\documentclass{rspublic}

\usepackage{graphics,amssymb,amsmath}

\begin{document}

\title[Advantage of a quantum over a classical player]{Advantage of a quantum player over
a classical one in 2x2 quantum games}
\author[A. P. Flitney and D. Abbott]{Adrian P. Flitney and Derek Abbott}
\affiliation{Centre for Biomedical Engineering (CBME) \\
and Department of Electrical and Electronic Engineering, \\
The University of Adelaide, Adelaide, South Australia 5005, Australia}
\label{firstpage}
\maketitle

\begin{abstract}{quantum games, game theory}
We study a general $2 \times 2$ symmetric, entangled, quantum game.
When one player has access only to classical strategies
while the other can use the full range of quantum strategies,
there are `miracle' moves available to the quantum player that
can direct the result of the game towards the quantum player's preferred result
regardless of the classical player's strategy.
The advantage pertaining to the quantum player is dependent
on the degree of entanglement.
Below a critical level,
dependent on the payoffs in the game,
the miracle move is of no advantage.
\end{abstract}

\section{Introduction}
Quantum game theory is an interesting new development in the fields
of game theory and quantum information.
First initiated by Meyer (1999),
a protocol for two player, two strategy ($2 \times 2$) games
was developed by Eisert and collaborators (1999)
and extended to multi-player games by Benjamin \& Hayden (2001$a$).
Where both players have access to the full set of quantum strategies
there is no Nash equilibrium (NE)
amongst pure strategies (Benjamin \& Hayden 2001$b$),
though there are an infinite set of equilibria
amongst mixed quantum strategies (Eisert \& Wilkens 2000).
A pure quantum strategy specifies a particular quantum operator to apply
contingent on the game situation,
whereas a mixed quantum strategy specifies
a probabilistic mixture of operators.
In a dynamical game one generally would not expect convergence to a NE.
In an entangled quantum game,
if the (pure) strategy of one player is known
the other player can produce any desired final state
by a suitable (pure) counter strategy,
assuring them of the maximum payoff.
Hence it is always possible for one of the players to improve his/her payoff
by a unilateral change in strategy.
For a discussion, see the
recent review of quantum games by Flitney \& Abbott (2002$a$).

When one player is restricted to classical moves
and the other is permitted the full quantum domain,
the quantum player has a clear advantage.
Eisert found that in a two player prisoners' dilemma
the quantum player could guarantee an expected payoff not less than that
of mutual cooperation,
while the classical player's reward was substantially smaller.
The advantage gained by the quantum player was found to be dependent
on the level of entanglement.
Below a critical level the quantum player could do no better than
adopting the classical dominant strategy.
It is interesting to speculate on the relationship between
the advantage obtainable by a quantum player over their classical rival
and the advantage a quantum algorithm has over a classical one.

In this work we extend the result of Eisert {\em et al} (1999)
and a later generalization by Du {\em et al} (2001$a,b$)
for prisoners' dilemma
to a general $2 \times 2$ quantum game.
Section~\ref{s-games} will summarize the protocol
for $2 \times 2$ entangled, quantum games.
In \S\ref{s-miracle} we determine the four different miracle moves,
depending on the game result most desired by the quantum player,
and consider the payoffs as a function of the degree of entanglement.
Section~\ref{s-critical} presents threshold values of the entanglement
for various game situations
and \S\ref{s-extensions} briefly considers extensions to larger strategic spaces.

\section{Quantum $2 \times 2$ games}
\label{s-games}
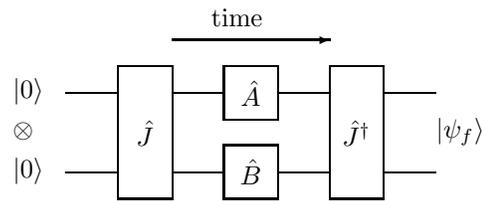
\begin{figure}
\begin{center}
\begin{picture}(180,100)(0,-10)
        \multiput(0,8)(0,30){2}{$|0\rangle$}
        \put(0,23){$\otimes$}
        \put(160,23){$|\psi_f\rangle$}

        \multiput(20,10)(40,0){4}{\line(1,0){20}}
        \multiput(20,40)(40,0){4}{\line(1,0){20}}
        \put(40,0){\framebox(20,50){$\hat{J}$}}
        \put(120,0){\framebox(20,50){$\hat{J}^{\dagger}$}}

        \put(80,30){\framebox(20,20){$\hat{A}$}}
        \put(80,0){\framebox(20,20){$\hat{B}$}}

        \put(60,60){\vector(1,0){60}}
        \put(75,65){time}
\end{picture}
\end{center}
\caption{The flow of information or qubits (solid lines)
in a general two person quantum game.
$\hat{A}$ is Alice's move, $\hat{B}$ is Bob's,
$\hat{J}$ is an entangling gate,
and $\hat{J}^{\dagger}$ is a disentangling gate.}
\label{f-game22}
\end{figure}

Figure~\ref{f-game22} is a protocol for a quantum game
between Alice and Bob.
The players' actions are encoded by qubits
that are initialized in the $|0\rangle$ state.
An entangling operator $\hat{J}$ is selected
which commutes with the direct product of any pair of classical strategies
utilized by the players.
Alice and Bob carry out local manipulations on their qubit by
unitary operators $\hat{A}$ and $\hat{B}$, respectively,
drawn from corresponding strategic spaces $S_A$ and $S_B$.
A projective measurement in the basis $\{|0\rangle, |1\rangle\}$
is carried out on the final state
and the payoffs are determined from the standard payoff matrix.
The final quantum state $|\psi_f \rangle$ is calculated by
\begin{equation}
| \psi_f \rangle = \hat{J}^{\dagger} (\hat{A} \otimes \hat{B}) \hat{J} \,
                        | \psi_i \rangle ,
\end{equation}
where
$|\psi_i \rangle = |00\rangle$ represents the initial state of the qubits.
The quantum game protocol contains the classical variant as a subset
since when $\hat{A}$ and $\hat{B}$ are classical operations
$\hat{J}^{\dagger}$ exactly cancels the effect of $\hat{J}$.
In the quantum game it is only
the expectation values of the players' payoffs that are important.
For Alice (Bob) we can write
\begin{equation}
\langle \$ \rangle = P_{00} |\langle \psi_f|00 \rangle|^2 +
                        P_{01} |\langle \psi_f|01 \rangle|^2 +
                        P_{10} |\langle \psi_f|10 \rangle|^2 +
                        P_{11} |\langle \psi_f|11 \rangle|^2
\end{equation}
where $P_{ij}$ is the payoff for Alice (Bob) associated with the game outcome
$ij; \; i, j \in \{0,1\}$.

The classical pure strategies correspond to
the identity operator $\hat{I}$
and the bit flip operator
\begin{equation}
\hat{F} \equiv \ri \hat{\sigma}_x = \left( \begin{array}{cc}
                                        0 & \ri \\
                                        \ri & 0
                                  \end{array} \right) .
\end{equation}
Without loss of generality,
an entangling operator $\hat{J}(\gamma)$
for an $N$-player game with two pure classical strategies
(an $N \times 2$ game)
may be written (Benjamin \& Hayden 2001$a$; Du {\em et al} 2001$a$)\footnote{
Any other choice of $\hat{J}$ would be equivalent,
via local unitary operations,
and would result only in a rotation of $|\psi_f\rangle$
in the complex plane,
consequently leading to the same game equilibria.}
\begin{equation}
\hat{J}(\gamma) = \exp \left( \ri \frac{\gamma}{2}
			\hat{\sigma}_x^{\otimes N} \right)
		= \hat{I}^{\otimes N} \cos \frac{\gamma}{2}
		+ \ri \hat{\sigma}_x^{\otimes N} \sin \frac{\gamma}{2} ,
\end{equation}
where $\gamma \in [0,\pi/2], \; \gamma = \pi/2$ corresponding to maximum
entanglement.
That is,
\begin{equation}
\hat{J}(\pi/2) |00 \ldots 0 \rangle =
	\frac{1}{\sqrt{2}}(|00 \ldots 0 \rangle + \ri |11 \ldots 1 \rangle) .
\end{equation}
A pure quantum strategy $\hat{U}(\theta, \alpha, \beta)$
is an SU(2) operator and may be written as
\begin{equation}
\label{e-su2}
\hat{U}(\theta, \alpha, \beta) =
        \left( \begin{array}{cc}
                e^{\ri \alpha} \cos (\theta/2) & \ri e^{\ri \beta} \sin (\theta/2) \\
                \ri e^{-\ri \beta} \sin (\theta/2) & e^{-\ri \alpha} \cos(\theta/2)
            \end{array} \right) ,
\end{equation}
where $\theta \in [0,\pi]$ and $\alpha, \beta \in [-\pi,\pi]$.
A classical mixed strategy can be simulated in the quantum game protocol
by an operator in the set $\tilde{U}(\theta) = \hat{U}(\theta,0,0)$.
Such a strategy corresponds to playing
$\hat{I}$ with probability $\cos^2 (\theta/2)$
and $\hat{F}$ with probability $\sin^2 (\theta/2)$.
Where both players use such strategies the game
is equivalent to the classical game.

\section{Miracle moves}
\label{s-miracle}
When both players have access to the full set of quantum operators,
for any $\hat{A} = \hat{U}(\theta,\alpha,\beta)$,
there exists $\hat{B} = \hat{U}(\theta, \alpha, -\frac{\pi}{2} - \beta)$,
such that
\begin{equation}
(\hat{A} \otimes \hat{I}) \frac{1}{\sqrt{2}} ( |00\rangle + \ri |11\rangle)
 = (\hat{I} \otimes \hat{B}) \frac{1}{\sqrt{2}} ( |00\rangle + \ri |11\rangle ) .
\end{equation}
That is, on the maximally entangled state,
any local unitary operation that Alice carries out on her qubit
is equivalent to a local unitary operation that Bob
carries out on his (Benjamin \& Hayden 2001$b$).
Hence either player can undo his/her opponent's move
(assuming it is known)
by choosing $\hat{U}(\theta, -\alpha, \frac{\pi}{2} - \beta)$
in response to $\hat{U}(\theta, \alpha, \beta)$.
Indeed, knowing the opponent's move,
either player can produce any desired final state.

We are interested in the classical-quantum game where one player, say Alice,
is restricted to $S_{cl} \equiv \{\tilde{U}(\theta) : \theta \in [0,\pi]\}$
while the other, Bob, has access to
$S_{q} \equiv \{\hat{U}(\theta,\alpha,\beta) :
	\theta \in [0,\pi]; \; \alpha, \beta \in [-\pi,\pi]\}$.
We shall refer to strategies in $S_{cl}$ as `classical'
in the sense that the player 
simply executes his/her two classical moves with fixed probabilities
and does not manipulate the phase of their qubit.
However, $\tilde{U}(\theta)$ only gives the same results
as a classical mixed strategy
when both players employ these strategies.
If Bob employs a quantum strategy he can
exploit the entanglement to his advantage. 
In this situation Bob has a distinct advantage since only he can produce
any desired final state by local operations on his qubit.
Without knowing Alice's move,
Bob's best plan is to play the `miracle' quantum move
consisting of assuming that Alice has played $\tilde{U}(\pi/2)$,
the `average' move from $S_{cl}$,
undoing this move by
\begin{equation}
\hat{V} = \hat{U}(\pi/2,0,\pi/2)
	= \frac{1}{\sqrt{2}}
		\left( \begin{array}{cc}
			1 & -1 \\
			1 & 1
		\end{array} \right) ,
\end{equation}
and then preparing his desired final state.
The operator
\begin{equation}
\hat{f} = \left( \begin{array}{cc}
                   0 & 1 \\
                   -1 & 0
          \end{array} \right)
\end{equation}
has the property
\begin{equation}
(\hat{I} \otimes \hat{f}) \frac{1}{\sqrt{2}} (|00\rangle + \ri |11\rangle)
	= (\hat{F} \otimes \hat{I}) \frac{1}{\sqrt{2}}
		(|00\rangle + \ri |11\rangle) ,
\end{equation}
so Bob can effectively flip Alice's qubit
as well as adjusting his own.

Suppose we have a general $2 \times 2$ game with payoffs
\begin{equation}
\begin{tabular}{c|ccc}
 & Bob: 0 & \hspace{5mm} & Bob: 1 \\
\hline
Alice: 0 & $(p,p')$ && $(q,q')$ \\
Alice: 1 & $(r,r')$ && $(s,s')$
\end{tabular}
\end{equation}
where the unprimed values refer to Alice's payoffs and the primed to Bob's.
Bob has four possible miracle moves
depending on the final state that he prefers:\footnote{In the published version of the paper
there is a mistake in the second line of Eq.~(\ref{eqn-miracle}). It should read
$\hat{M}_{01} = \hat{F} \hat{V} = \frac{i}{\sqrt{2}} \begin{pmatrix} 1 & 1 \\ 1 & -1 \end{pmatrix}$.}
\begin{equation}
\label{eqn-miracle}
\begin{split}
\hat{M}_{00} &= \hat{V} , \\
\hat{M}_{01} &= \hat{F} \hat{V}
	      = \frac{\ri}{\sqrt{2}}
		\left( \begin{array}{cc} 1 & 1 \\ 1 & -1 \end{array} \right) , \\
\hat{M}_{10} &= \hat{f} \hat{V}
	      = \frac{1}{\sqrt{2}}
		\left( \begin{array}{cc} 1 & 1 \\ -1 & 1 \end{array} \right) , \\
\hat{M}_{11} &= \hat{F} \hat{f} \hat{V}
	      = \frac{\ri}{\sqrt{2}}
		\left( \begin{array}{cc} -1 & 1 \\ 1 & 1 \end{array} \right) ,
\end{split}
\end{equation}
given a preference for $|00\rangle$, $|01\rangle$, $|10\rangle$, or
$|11\rangle$, respectively.
In the absence of entanglement,
any $\hat{M}_{ij}$ is equivalent to $\tilde{U}(\pi/2)$,
that is, the mixed classical strategy of flipping or not-flipping
with equal probability.

When we use an entangling operator
$\hat{J}(\gamma)$ for an arbitrary $\gamma \in [0,\pi/2]$,
the expectation value of Alice's payoff
if she plays $\tilde{U}(\theta)$ against Bob's miracle moves are, respectively,
\begin{equation}
\label{e-payoffs}
\begin{split}
\langle \$_{00} \rangle
&= \frac{p}{2} (\cos \frac{\theta}{2} + \sin \frac{\theta}{2} \sin \gamma)^2
	\,+\, \frac{q}{2} \cos^2 \frac{\theta}{2} \cos^2 \gamma \\
& \makebox[2cm]{}
	\,+\, \frac{r}{2} (\sin \frac{\theta}{2} - \cos \frac{\theta}{2} \sin \gamma)^2
	\,+\, \frac{s}{2} \sin^2 \frac{\theta}{2} \cos^2 \gamma , \\
\langle \$_{01} \rangle
&= \frac{p}{2} \cos^2 \frac{\theta}{2} \cos^2 \gamma
	\,+\, \frac{q}{2} (\cos \frac{\theta}{2} + \sin \frac{\theta}{2} \sin \gamma)^2
	\,+\, \frac{r}{2} \sin^2 \frac{\theta}{2} \cos^2 \gamma \\
& \makebox[2cm]{}
	\,+\, \frac{s}{2} (\sin \frac{\theta}{2} -
		\cos \frac{\theta}{2} \sin \gamma)^2 , \\
\langle \$_{10} \rangle
&= \frac{p}{2} (\cos \frac{\theta}{2} - \sin \frac{\theta}{2} \sin \gamma)^2
	\,+\, \frac{q}{2} \cos^2 \frac{\theta}{2} \cos^2 \gamma \\
& \makebox[2cm]{}
	\,+\, \frac{r}{2} (\sin \frac{\theta}{2} + \cos \frac{\theta}{2} \sin \gamma)^2
	\,+\, \frac{s}{2} \sin^2 \frac{\theta}{2} \cos^2 \gamma , \\
\langle \$_{11} \rangle
&= \frac{p}{2} \cos^2 \frac{\theta}{2} \cos^2 \gamma
	\,+\, \frac{q}{2} (\cos \frac{\theta}{2} - \sin \frac{\theta}{2} \sin \gamma)^2
	\,+\, \frac{r}{2} \sin^2 \frac{\theta}{2} \cos^2 \gamma
\\
& \makebox[2cm]{}
	\,+\, \frac{s}{2} (\sin \frac{\theta}{2} + \cos \frac{\theta}{2}
		\sin \gamma)^2.
\end{split}
\end{equation}
We add primes to $p,q,r,$ and $s$ to get Bob's payoffs.
Although the miracle moves are in some sense best for Bob,
in that they guarantee a certain minimum payoff against any classical strategy
from Alice, there is not necessarily any NE amongst pure strategies
in the classical-quantum game.

\section{Critical entanglements}
\label{s-critical}
In each of the four cases of equation~(\ref{e-payoffs})
there can be critical values
of the entanglement parameter $\gamma$ below which the quantum player no longer
has an advantage.
We will consider some examples.
The most interesting games are those that pose some sort of dilemma
for the players.
A non-technical discussion of various dilemmas in $2 \times 2$ game theory
is given in Poundstone (1992)
from which we have taken the names of the following games.
The results for prisoners' dilemma,
using the standard payoffs for that game,
were found by Eisert {\em et al} (1999)
and, for generalized payoffs, by Du {\em et al} (2001$a$).

Below we introduce a number of games,
discuss the dilemma faced by the players
and their possible strategies.
The games, along with some important equilibria,
are summarized in table~\ref{t-games}.
Detailed results for the various threshold values of the entanglement parameter
are given for the game of chicken.
A summary of the thresholds for the collection of games
is given in table~\ref{t-results}.
In the following, the payoffs shall be designated $a,b,c,$ and $d$ with
$a > b > c > d$.
The two pure classical strategies for the players
are referred to as cooperation ($C$) and defection ($D$),
for reasons that shall soon become apparent.
The NE's referred to are in classical pure strategies
unless otherwise indicated.

\begin{table}
\begin{tabular}{lccccc}
\hline
game & payoff matrix & NE payoffs & PO payoffs & condition &
	$(a,b,c,d)$ \\
\hline
chicken
	& $ \begin{array}{cc} (b,b) & (c,a) \\ (a,c) & (d,d) \end{array} $
	& $(a,c)$ or $(c,a)$ & $(b,b)$ & $2 b > a + c$ &
	$(4,3,1,0)$ \\
&&&&& \\
PD
	& $ \begin{array}{cc} (b,b) & (d,a) \\ (a,d) & (c,c) \end{array} $
	& $(c,c)$ & $(b,b)$ & $2 b > a + d$ & $(5,3,1,0)$ \\
&&&&& \\
deadlock
	& $ \begin{array}{cc} (c,c) & (d,a) \\ (a,d) & (b,b) \end{array} $
 	& $(b,b)$ & $(b,b)$ & $2 b > a + d$ & $(3,2,1,0)$ \\
&&&&& \\
stag hunt
	& $ \begin{array}{cc} (a,a) & (d,b) \\ (b,d) & (c,c) \end{array} $
	& $(a,a)$ or $(c,c)$ & $(a,a)$ & & $(3,2,1,0)$ \\
&&&&& \\
BoS
	& $ \begin{array}{cc} (a,b) & (c,c) \\ (c,c) & (b,a) \end{array} $
	& $(a,b)$ or $(b,a)$ & $(a,b)$ or $(b,a)$ & &
	$(2,1,0)$ \\
\hline
\end{tabular}
\caption{Payoff matrices for some $2 \times 2$ games}
\longcaption{A summary of payoff matrices
with NE and PO results for various classical games.
PD is the prisoners' dilemma and BoS is the battle of the sexes.
In the matrices, the rows correspond to Alices's options
of cooperation ($C$) and defection ($D$), respectively,
while the columns are likewise for Bob's.
In the parentheses, the first payoff is Alice's, the second Bob's
and $a > b > c > d$.
The condition specifies a constraint on the values of $a,b,c,$ and $d$
necessary to create the dilemma.
The final column gives standard values for the payoffs.}
\label{t-games}
\end{table}

\subsection{Chicken}
The archetypal version of chicken is described as follows:
\begin{quote}
The two players are driving towards each other along the centre of an empty road.
Their possible actions are to swerve at the last minute (cooperate)
or not to swerve (defect).
If only one player swerves he/she is the `chicken'
and gets a poor payoff,
while the other player is the `hero' and scores best.
If both swerve they get an intermediate result
but clearly the worst possible scenario is for neither player to swerve.
\end{quote}
Such a situation often arises in the military/diplomatic posturing amongst nations.
Each does best if the other backs down against their strong stance,
but the mutual worst result is to go to war!
The situation is described by the payoff matrix
\begin{equation}
\begin{tabular}{c|ccc}
 & Bob: $C$ & \hspace{5mm} & Bob: $D$ \\
\hline
Alice: $C$ & $(b,b)$ && $(c,a)$ \\
Alice: $D$ & $(a,c)$ && $(d,d)$
\end{tabular}
\end{equation}
Provided $2b > a+c$, the Pareto optimal (PO) result,
the one for which it is not possible to improve the payoff of one player
without reducing the payoff of the other,
is mutual cooperation.
In the discussion below
we shall choose $(a,b,c,d) = (4,3,1,0)$,
satisfying this condition,
whenever we want a numerical example of the payoffs.
There are two NE in the classical game, $CD$ and $DC$,
from which neither player can improve their result by a unilateral change
in strategy.
Hence the rational player hypothesised by game theory is faced with a dilemma
for which there is no solution:
the game is symmetric yet both players want to do the opposite
of the other.
For the chosen set of numerical payoffs
there is a unique NE in mixed classical strategies:
each player cooperates or defects with probability 1/2.
In our protocol this corresponds to both players selecting $\tilde{U}(\pi/2)$.

A quantum version of chicken has been discussed in the literature
(Marinatto \& Weber 2000$a,b$; Benjamin 2000).
In this, a final state of a player's qubit being $|0\rangle$
corresponds to the player having cooperated,
while $|1\rangle$ corresponds to having defected.

The preferred outcome for Bob is $CD$ or $|01\rangle$,
so he will play $\hat{M}_{01}$.
If Alice cooperates, the payoffs are
\begin{equation}
\begin{split}
\langle \$_A \rangle &= \frac{b-d}{2} \cos^2 \gamma \,+\, \frac{c+d}{2} , \\
\langle \$_B \rangle &= \frac{b-d}{2} \cos^2 \gamma \,+\, \frac{a+d}{2} .
\end{split}
\end{equation}
Increasing entanglement is bad for the both players.
However, Bob out scores Alice by $(a-c)/2$ for all entanglements
and does better than the poorer of his two NE results ($c$) provided
\begin{equation}
\sin \gamma < \sqrt{\frac{a+b-2c}{b-d}}
\end{equation}
which, for the payoffs (4,3,1,0), means that $\gamma$ can take any value.
He performs better than the mutual cooperation result ($b$) provided
\begin{equation}
\sin \gamma < \sqrt{\frac{a-b}{b-d}}
\end{equation}
which yields a value of $1/\sqrt{3}$ for the chosen payoffs.

Suppose instead that Alice defects.
The payoffs are now
\begin{equation}
\begin{split}
\langle \$_A \rangle &= \frac{a-c}{2} \cos^2 \gamma \,+\, \frac{c+d}{2} , \\
\langle \$_B \rangle &= \frac{a-c}{2} \sin^2 \gamma \,+\, \frac{c+d}{2} .
\end{split}
\end{equation}
Increasing entanglement improves Bob's result
and worsens Alice's.
Bob scores better than Alice provided $\gamma > \pi/4$,
regardless of the numerical value of the payoffs.
Bob does better than his worst NE result ($c$) when
\begin{equation}
\label{e-thresh}
\sin \gamma > \sqrt{\frac{c-d}{a-c}} \;,
\end{equation}
which yields a value of $1/\sqrt{3}$ for the default payoffs,
and better than his PO result ($b$) when
\begin{equation}
\sin \gamma > \sqrt{\frac{2b-c-d}{a-c}} \;,
\end{equation}
which has no solution for the default values.
Thus, except for specially adjusted values of the payoffs,
Bob cannot assure himself of a payoff
at least as good as that achievable by mutual cooperation.
However, Bob escapes from his dilemma for a sufficient degree of entanglement
as follows.
Against $\hat{M}_{01}$,
Alice's optimal strategy from the set $S_{cl}$ is given by
\begin{equation}
\tan \theta = \frac{2(c-d)}{b+c-a-d} \: \frac{\sin \gamma}{\cos^2 \gamma} \;.
\end{equation}
For $(a,b,c,d) = (4,3,1,0)$
this gives $\theta = \pi/2$.
Since $\hat{M}_{01}$ is Bob's best counter to $\tilde{U}(\pi/2)$
these strategies form a NE in classical-quantum game of chicken
and are the preferred strategies of the players.
For this choice,
above an entanglement of $\gamma = \pi/6$,
Bob performs better than his mutual cooperation result.

\begin{figure}
\includegraphics{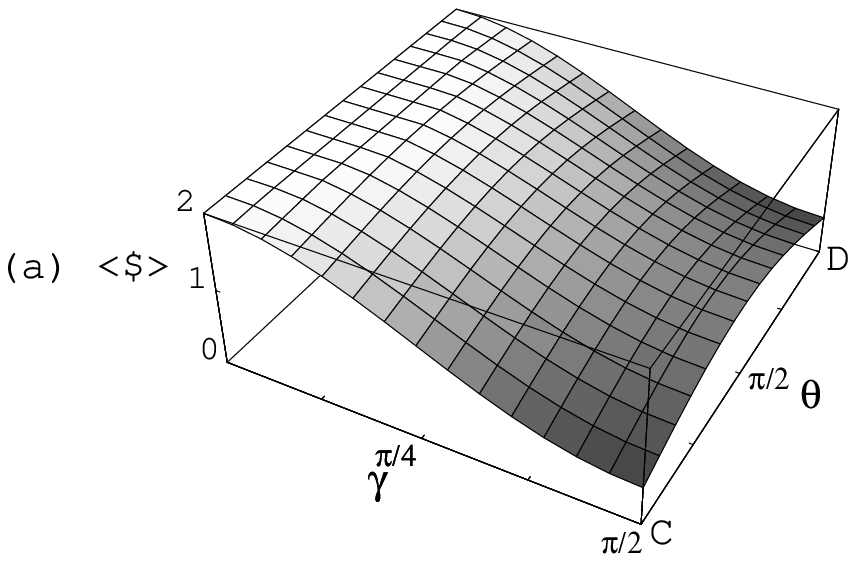}
\includegraphics{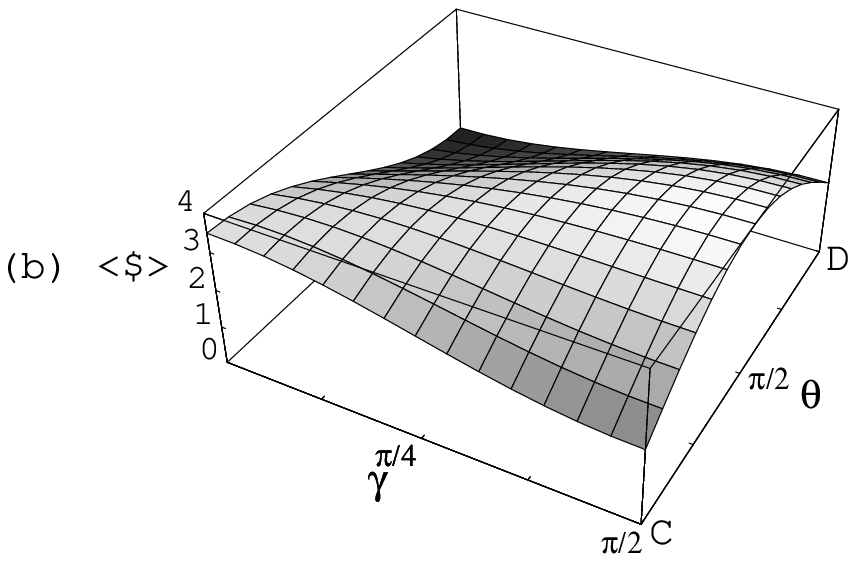}
\caption{In the game of chicken,
the expected payoffs for (a) Alice and (b) Bob
when Bob plays the quantum miracle move $\hat{M}_{01}$,
as a function of Alice's strategy
($\theta = 0$ corresponds to cooperation
and $\theta =\pi$ corresponds to defection)
and the degree of entanglement $\gamma$.
The surfaces are drawn for payoffs $(a,b,c,d) = (4,3,1,0)$.
If Alice knows that Bob is going to play the quantum miracle move,
she does best by choosing the crest of the curve, $\theta=\pi/2$,
irrespective of the level of entanglement.
Against this strategy
Bob scores between two and four,
an improvement for all $\gamma > 0$
over the payoff he could expect playing a classical strategy.}
\label{f-chicken}
\end{figure}

\begin{figure}
\includegraphics{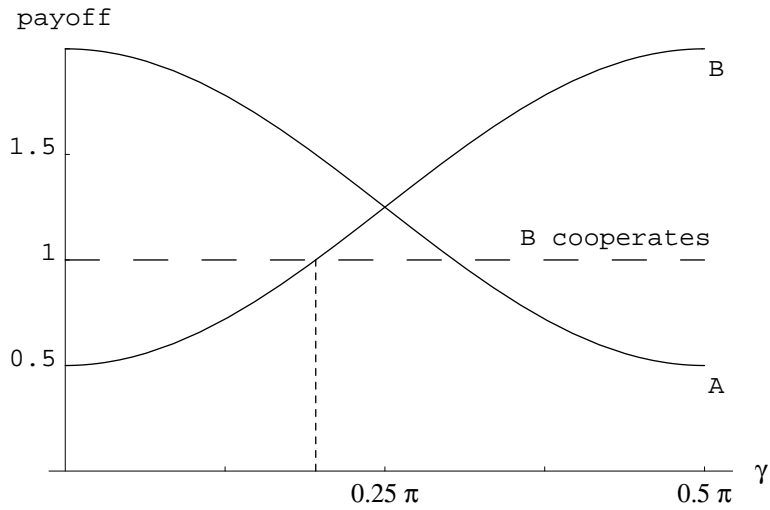}
\caption{The payoffs for Alice and Bob
versus the level of entanglement ($\gamma$)
in the game of chicken
when Alice defects.
The standard payoffs $(a,b,c,d) = (4,3,1,0)$ are chosen.
The solid lines correspond to the results
when Bob plays the quantum move $\hat{M}_{01}$
and the dashed line gives Bob's payoff when
he cooperates.
Below an entanglement of $\arcsin (1/\sqrt{3})$
(short dashes)
Bob does best,
against a defecting Alice,
by switching to the strategy `always cooperate.'}
\label{f-chickend}
\end{figure}

The expected payoffs for Alice and Bob as a function of Alice's strategy
and the degree of entanglement
are shown in figure~\ref{f-chicken}.
In figure~\ref{f-chickend} we can see that if Bob wishes
to maximize the minimum payoff he receives,
he should alter his strategy from the quantum move $\hat{M}_{01}$
to cooperation,
once the entanglement drops below $\arcsin (1/\sqrt{3})$.

\subsection{Prisoners' dilemma}
The most famous dilemma is the prisoners' dilemma.
This may be specified in general as
\begin{equation}
\begin{tabular}{c|ccc}
 & Bob: C & \hspace{5mm} & Bob: D \\
\hline
Alice: C & $(b,b)$ && $(d,a)$ \\
Alice: D & $(a,d)$ && $(c,c)$
\end{tabular}
\end{equation}
In the classical game,
the strategy `always defect' dominates since it gives a better payoff
than cooperation against any strategy by the opponent.
Hence, the NE for the prisoners' dilemma is mutual defection,
resulting in a payoff of $c$ to both players.
However, both players would have done better with mutual cooperation,
resulting in a payoff of $b$,
giving rise to a dilemma that occurs in many social and political situations.
The sizes of the payoffs are generally adjusted so that $2b > a+d$
making mutual cooperation the PO outcome.
The most common set of payoffs is
$(a,b,c,d) = (5,3,1,0)$.

In the classical-quantum game
Bob can help engineer his preferred result of $|01\rangle$ ($CD$)
by adopting the strategy $\hat{M}_{01}$.
The most important critical value of the entanglement parameter $\gamma$
is the threshold below which Bob performs worse with his miracle move
than he would if he chose the classical dominant strategy of `always defect'.
This occurs for
\begin{equation}
\label{e-pdthresh}
\sin \gamma = \sqrt{\frac{c-d}{a-d}} \;,
\end{equation}
which yields the value $\sqrt{1/5}$ for the usual payoffs.
As noted in Du {\em et al} (2001$b$),
below this level of entanglement the quantum version of prisoners' dilemma
behaves classically with a NE of mutual defection.

\subsection{Deadlock}
Deadlock is characterized by reversing
the payoffs for mutual cooperation and defection
in the prisoners' dilemma:
\begin{equation}
\begin{tabular}{c|ccc}
 & Bob: $C$ & \hspace{5mm} & Bob: $D$ \\
\hline
Alice: $C$ & $(c,c)$ && $(d,a)$ \\
Alice: $D$ & $(a,d)$ && $(b,b)$
\end{tabular}
\end{equation}
Defection is again the dominant strategy and
there is even less incentive for the players to cooperate in this game
than in the prisoners' dilemma since the PO result
is mutual defection.
However, both players would prefer if their opponent cooperated 
so they could stab them in the back by defecting and achieve the maximum payoff
of $a$.
There is no advantage to cooperating so there is no real dilemma
in the classical game.
In the classical-quantum game Bob can again use his quantum skills
to engineer at least partial cooperation from Alice,
against any possible strategy from her,
by playing $\hat{M}_{01}$.

\subsection{Stag hunt}
In stag hunt both players prefer the outcome of mutual cooperation
since it gives a payoff superior to all other outcomes.
However, each are afraid of defection by the other
which, although it reduces the defecting player's payoff,
has a more detrimental effect on the cooperator's payoff,
as indicated in the payoff matrix below:
\begin{equation}
\begin{tabular}{c|ccc}
 & Bob: $C$ & \hspace{5mm} & Bob: $D$ \\
\hline
Alice: $C$ & $(a,a)$ && $(d,b)$ \\
Alice: $D$ & $(b,d)$ && $(c,c)$
\end{tabular}
\end{equation}
Both mutual cooperation and mutual defection are NE,
but the former is the PO result.
There is no dilemma when two rational players meet.
Both recognize the preferred result and have no reason,
given their recognition of the rationality of the other player,
to defect.
Mutual defection will result only if both players
allow fear to dominate over rationality.
This situation is not changed in the classical-quantum game.
However, having the ability to play quantum moves may be of advantage
when the classical player is irrational
in the sense that they do not try to maximize their own payoff.
In that case the quantum player should choose to play the strategy
$\hat{M}_{00}$
to steer the result towards the mutual cooperation outcome.

\subsection{Battle of the sexes}
In this game Alice and Bob each want the company of the other
in some activity,
but their preferred activity differs:
opera ($O$) for Alice and television ($T$) for Bob.
If the players end up doing different activities
both are punished by a poor payoff.
In matrix form this game can be represented as
\begin{equation}
\begin{tabular}{c|ccc}
 & Bob: $O$ & \hspace{5mm} & Bob: $T$ \\
\hline
Alice: $O$ & $(a,b)$ && $(c,c)$ \\
Alice: $T$ & $(c,c)$ && $(b,a)$
\end{tabular}
\end{equation}
The options on the main diagonal are both PO and are NE
but there is no clear way of deciding between them.
Bob's quantum strategy will be to choose $\hat{M}_{11}$ to steer the game
towards his preferred result of $TT$.
Marinatto and Weber (2000$a,b$; Benjamin 2000)
discuss a quantum version of battle of the sexes
using a slightly different protocol for a quantum game than the one used
in the current work.

\begin{figure}
\includegraphics{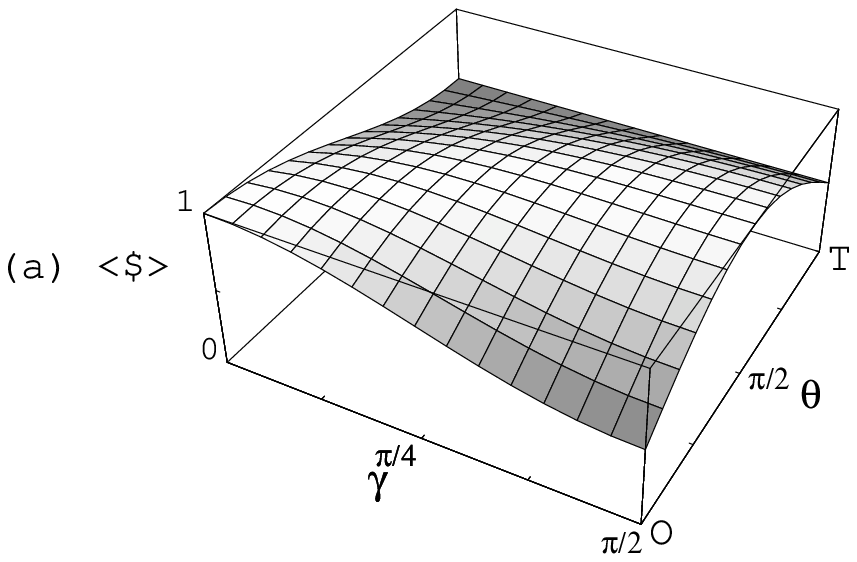}
\includegraphics{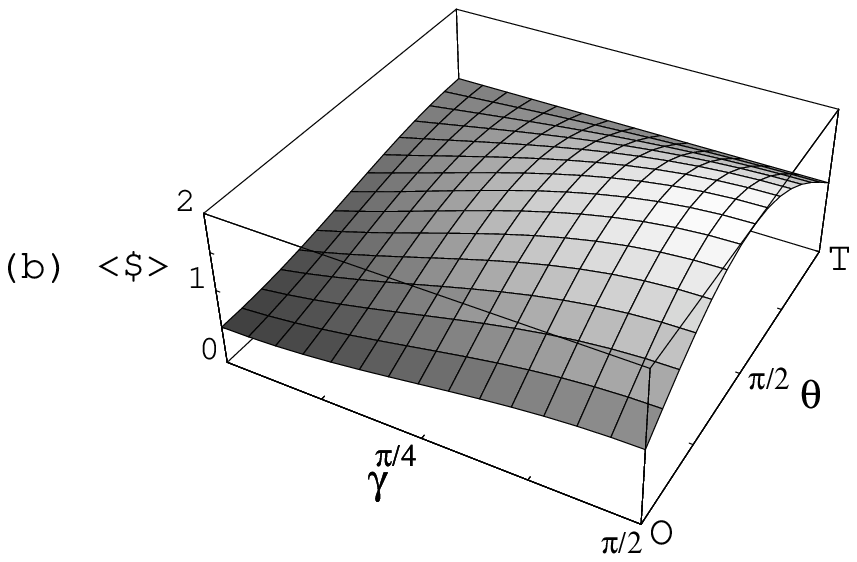}
\caption{In the battle of the sexes,
the expected payoffs for (a) Alice and (b) Bob
when Bob plays the quantum miracle move $\hat{M}_{11}$,
as a function of Alice's strategy
($\theta = 0$ corresponds to opera
and $\theta =\pi$ corresponds to television)
and the degree of entanglement $\gamma$.
The surfaces are drawn for payoffs $(a,b,c) = (2,1,0)$.
If Alice knows that Bob is going to play the quantum miracle move,
she does best by choosing the crest of the curve,
so her optimal strategy changes from $O$ for no entanglement,
to $\theta = \pi/2$ for full entanglement.
Against this strategy,
Bob starts to score better than one
for an entanglement exceeding approximately $\pi/5$.}
\label{f-BoS}
\end{figure}

With $\hat{M}_{11}$, Bob out scores Alice provided $\gamma > \pi/4$,
but is only assured of scoring at least as well as the poorer of his two NE
results ($b$) for full entanglement,
and is never certain of bettering it.
The quantum move, however, is better than using a fair coin to decide between
$\hat{O}$ and $\hat{T}$ for $\gamma > 0$,
and equivalent to it for $\gamma = 0$.
Hence, even though Bob cannot be assured of scoring greater than $b$
he can improve his worst case payoff for any $\gamma > 0$.
Figure~\ref{f-BoS} shows Alice and Bob's payoffs
as a function of the degree of entanglement and Alice's strategy.

\begin{table}
\begin{tabular}{lcccc}
\hline
game & strategies &
$\langle \$_B \rangle > \langle \$_{A} \rangle$ &
$\langle \$_B \rangle > \$_B$ (NE) &
$\langle \$_B \rangle > \$_B$ (PO) \\
\hline
chicken
& $\hat{C}$, $\hat{M}_{01}$ &
always & $< (a+b-2c)/(b-d)$ & $< (a-b)/(b-d)$ \\
& $\hat{D}$, $\hat{M}_{01}$ &
$1/2$ & $(c-d)/(a-c)$ & $(2b-c-d)/(a-c)$ \\
&&&& \\
PD
& $\hat{C}$, $\hat{M}_{01}$ &
always & always & $(a-b)/(c-d)$ \\
& $\hat{D}$, $\hat{M}_{01}$ & $d/(2(a-d))$ &
$(c-d)/(a-d)$ & $(2b-c-d)/(a-d)$ \\
&&&& \\
deadlock
& $\hat{C}$, $\hat{M}_{01}$ &
always & $(2b-a-c)/(b-c)$ & $(2b-a-c)/(b-c)$ \\
& $\hat{D}$, $\hat{M}_{01}$ &
$1/2$ & $(b-d)/(a-d)$ & $(b-d)/(a-d)$ \\
&&&& \\
stag hunt
& $\hat{C}$, $\hat{M}_{00}$ &
$< 1/2$ & $(c-d)/(a-c)$ & never \\
& $\hat{D}$, $\hat{M}_{00}$ &
never & $ < (a+b-2c)/(b-d)$ & never \\
&&&& \\
BoS
& $\hat{O}$, $\hat{M}_{11}$ &
$1/2$ & $(b-c)/(a-b)$ & $(b-c)/(a-b)$ \\
& $\hat{T}$, $\hat{M}_{11}$ &
always & if $a+c > 2b$ & if $a+c > 2b$ \\
\hline
\end{tabular}
\caption{Critical entanglements for $2 \times 2$ quantum games.}
\longcaption{Values of $\sin^2 \gamma$ above which
(or below which where indicated by `$<$') the expected value of Bob's
payoff exceeds, respectively, Alice's payoff,
Bob's classical NE payoff
and, Bob's payoff for the PO outcome.
Where there are two NE (or PO) results,
the one where Bob's payoff is smallest is used.
The strategies are Alice's and Bob's, respectively.
In the last line,
`if $a + c > 2 b$' refers to a condition on the numerical values of the payoffs
and not to a condition on $\gamma$.}
\label{t-results}
\end{table}

\section{Extensions}
\label{s-extensions}
The situation is more complex for multi-player games.
No longer can a quantum player playing against classical ones
engineer any desired final state,
even if the opponents' moves are known.
However, a player can never be worse by having access to the quantum domain
since this includes the classical possibilities as a subset.

In two player games with more than two pure classical strategies
the prospects for the quantum player are better.
Some entangled, quantum $2 \times 3$ games have been considered
in the literature (Iqbal \& Toor 2001; Flitney \& Abbott 2002$b$).
Here the full set of quantum strategies is SU(3)
and there are nine possible miracle moves (before considering symmetries).
$S_{cl}$, the strategies that do not manipulate the phase of the player's
qutrit (a qutrit is the three state equivalent of a qubit)
can be written as the product of three rotations,
each parameterized by a rotation angle.
Since the form is not unique,
it is much more difficult to say what constitutes the average move from this set,
so the expressions for the miracle moves are open to debate.
Also, an entangling operator for a general $2 \times 3$ quantum game
has not been given in the literature.
Nevertheless, the quantum player will still be able to manipulate the
result of the game to increase the probability of his/her favoured result.

\section{Conclusion}
With a sufficient degree of entanglement,
the quantum player in a classical-quantum two player game can use
the extra possibilities available to help
steer the game towards their most desired result,
giving a payoff above that achievable
by classical strategies alone.
We have given the four miracle moves in quantum $2 \times 2$ game theory
and show when they can be of use in several game theory dilemmas.
There are critical values of the entanglement parameter $\gamma$
below (or occasionally above) which it is no longer
an advantage to have access to quantum moves.
That is, where the quantum player
can no longer outscore his/her classical Nash equilibrium result.
These represent a phase change in the classical-quantum game
where a switch between the quantum miracle move
and the dominant classical strategy is warranted.
With typical values for the payoffs
and a classical player opting for his/her best strategy,
the critical value for $\sin \gamma$ is
$\sqrt{1/3}$ for chicken,
$\sqrt{1/5}$ for prisoners' dilemma
and $\sqrt{2/3}$ for deadlock,
while for stag hunt
there is no particular advantage to the quantum player.
In the battle of the sexes there is no clear threshold
but for any non-zero entanglement Bob can improve his worst case result.

The quantum player's advantage is not as strong
in classical-quantum multi-player games
but in multi-strategy, two player games,
depending on the level of entanglement,
the quantum player would again have access to moves that improve his/her result.
The calculation of these moves is problematic because of the larger number
of degrees of freedom and has not been attempted here.

\begin{acknowledgements}
This work was supported by GTECH Corporation Australia
with the assistance of the SA Lotteries Commission (Australia).
\end{acknowledgements}

\vfill

\pagebreak

\label{lastpage}
\end{document}